\documentclass[12pt]{article}
\usepackage{amsmath}
\usepackage{amssymb}
\usepackage{amsfonts}
\usepackage{amscd}
\usepackage{delarray}
\usepackage{graphicx}

\usepackage{amsmath,amsmath,amsfonts,amssymb,color,bbm}

\usepackage{graphicx}

\def\Symp#1,#2,#3,#4.{\left[\left(\begin{array}{c}#1\\#2\end{array}\right),\left(\begin{array}{c}#3\\#4\end{array}\right)\right]}
\def\Vec#1,#2.{\left(\!\begin{array}{c}#1\\#2\end{array}\!\right)}
\def\vec#1,#2.{{#1\choose{#2}}}
\newcommand{\ket} [1] {\vert #1 \rangle}

\newcommand{\beq}{\begin{equation}}
\newcommand{\eeq}{\end{equation}}
\newcommand{\beqa}{\begin{eqnarray}}
\newcommand{\eeqa}{\end{eqnarray}}

\begin{document}

\title{Experimental proposal for testing the Emergence of Environment Induced (EIN) Classical Selection rules with Biological Systems.}
\date{}
\author{}
\maketitle
\vglue -1.8truecm
\centerline{Thomas Durt\footnote{TONA Vrije Universiteit Brussel, Pleinlaan 2, B-1050
Brussels, Belgium. \texttt{email:
thomdurt@vub.ac.be}} }

%



\begin{abstract}
According to the so-called  \textit{Quantum Darwinist} approach, the emergence of  \textit{``classical islands''} from a quantum background is assumed to obey a (selection) principle of maximal information. We illustrate this idea by considering the coupling of two oscillators (modes). As our approach suggests that the classical limit could have emerged throughout  a long and progressive Evolution mechanism, it is likely that primitive living organisms behave in a ``more quantum'', ``less classical'' way than more evolved ones.
This brings us to seriously  consider the possibility to measure departures from classicality exhibited by biological systems.  We describe an experimental proposal the aimed at revealing the presence of entanglement in the biophotonic radiation emitted by biological sources.
\end{abstract}


\maketitle

\section*{Introduction and overview}
The picture of the world elaborated by mathematicians and  physicists such as Laplace and Maxwell, at the golden age of classical physics, possessed a very appealing property: solidity.  Physical phenomena were, in the monolithic world view shared by the overwhelming majority of the scientists of that period, supposed to obey deterministic rules, in an arena constituted by Newtonian space-time (that is, Euclidean three-dimensional space supplemented by an external, absolute, time). The mathematical and geometrical rules that governed the whole construction were supposed to obey the rules of classical logics. It is only later that the foundational aspects of this comforting and comfortable representation of the world have been questioned, throughout several crises. 

In physics, the two main crises coincided with the development of modern relativity theories by Poincar\'e and Einstein and with the advent of the quantum theory. In mathematics, they  correspond to the development of non-Euclidean geometries, but also of non-classical logics, to the questioning of the concept of number and to foundational debates about the nature of the continuum.  

In a previous paper \cite{QD} we investigated the so-called Quantum Darwinist approach \cite{reviewzurek,wikipedia,EIN} according to which the emergence  of ``classicality'' could obey a 
(selection) principle of maximal information. 
Our analysis suggested that our classical preconceptions about the world could indeed possibly have been elaborated throughout a slow evolutionary process, according to a principle of maximization of the relevant information.

Considered so, classical physics would be an idealized limit-case, the visible part of an iceberg of which the foundations could obey radically non-classical rules (as for instance quantum mechanics).

The goal of the present paper is to investigate in which circumstances departures from these classical situations, if they exist, are likely to be observed, and to propose experimental schemes that would allow us to observe these departures. As we shall discuss, a fine study of the correlations exhibited by photons produced during the so-called long range delayed biophotonic emission process \cite{biowiki,gurwitsch,strehler,popp81}, an ultra weak optical radiation of biological origin, could possibly mark a departure from classical paradigms and reveal a specifically quantum signature.

In the first section we outline the general principles of the Quantum Darwinist approach, in which the emergence of ``classical islands'' is assumed to obey a principle of maximal information (this is the essence of the Environment Induced Selection rule expressed by Zurek and coworkers in the Decoherence approach to Quantum Mechanics \cite{decoherence}). In the Quantum Darwinist approach (a fine structure of the Decoherence Program \cite{wikipedia}), it is assumed that this principle of maximal information acts as a Darwinist selection principle that, throughout eons of evolution, privileged  classical representations (or pictures) of the world (that correspond to the ``classical, decoherence-free preferred bases''). In the following we shall take for granted that these classical islands correspond grosso modo to the parts of the quantum world that remain in a separable state throughout their evolution\footnote{Classicality is considered here in a very restricted sense: we say that a representation of two quantum systems $A$ and $B$ is classical whenever it makes sense to assume that $A$ and $B$ remain separable throughout their interaction. This means, in the case that the quantum state of $A$ and $B$ is pure, that it remains factorisable throughout time. It is out of the scope of the present paper to investigate the relations between classical islands, decoherence, and classical LOGICS but it is worth mentioning that this program has been partially completed in the framework of the {\it Consistent History Approach} \cite{omnes,griffith}.}.

In the second section, we illustrate these concepts by a very simple example: two quantum oscillators interact without getting entangled with each other (in other words they remain in a separable, so-called factorisable state throughout their interaction). We show that this occurs when their states are so-called coherent states. These states, sometimes called quasi-classical states, minimize uncertainty relations, and they can be considered to provide the most fidel quantum realisation of classical phase-space states in the case of harmonic oscillators.

If the Quantum Darwinist approach is valid, it is very likely that primitive organisms like cells and bacteria exhibit non-classical features (footnote 9 of ref.\cite{QD}). This is because they would be less constrained by environment induced selection rules, due to the fact that they are ``archaic'' forms of life\footnote{Besides, it is also likely that they are less constrained by a principle of maximal information due to the fact that they do not possess the nervous system necessary for treating and integrating information in a sophisticated manner.}. Considered so, one should expect that quantum weirdness would play a non-negligible role during the biological activies of these organisms. In the third section, we present certain experimental facts, recently gathered, that could indicate that so-called Light Harvesting Bacteria \cite{fleming} exhibit a high level of quantum coherence. We also emit the hypothesis that a still partially ununderstood phenomenon, the so-called Long Term Delayed Biological Fluorescence  \cite{biowiki} could also exhibit interesting, non-trivial, quantum features and propose an experimental procedure that would allow us to test this hypothesis.
  
  \section{Quantum Darwinism and the Classical Cartesian Picture.} 
Presently, it is still an open question to know whether quantum mechanics is necessary
 in order to describe the way that our brain functions. It is even an open question to know whether the non-classical aspects of quantum mechanics play a fundamental role in biological processes at all. It is for instance an open question to know whether or not quantum coherence must be invoked in
order to explain intra-cellular processes\footnote{Nothing illustrates better the present situation than this
quote of Eisert and Wiseman\cite{[5]}:
``When you have excluded the trivial, whatever remains, however improbable, must
be a good topic for a debate''...}.

Nevertheless, quantum mechanics appears to be astonishingly adequate in order to describe the
 material world in which we live. It is therefore interesting to investigate to which extent quantum physics plays a role in biological processes, and, more, to ask whether it could be that biological processes exhibit some form of quantum weirdness. A key concept in these issues is the so-called quantum entanglement.

  The term entanglement was first
introduced by Schr\"odinger who described it as the characteristic trait of quantum mechanics,
``the one that enforces its entire departure from classical
lines of thought'' \cite{erwin}. Bell's inequalities \cite{key-73} show that when two systems are prepared in an entangled
state, the knowledge of the whole cannot be reduced to the knowledge of the parts, and that to
some extent the systems lose their individuality. It is only when systems are not entangled that they behave as separable systems. It can be shown that whenever two distant systems are
in an entangled (pure) state, there exist well-chosen observables such that the associated
correlations do not admit a local realist explanation \cite{laudisa}, which is revealed by the violation of
well-chosen Bell's inequalities \cite{34}. Considered so, entanglement reintroduces holism\footnote{Holism is a rather vague concept that possesses several definitions, often mutually exclusive \cite{seevinck}. Here we mean that the quantum theory is holistic in the sense there can be a relevant difference in the whole without a difference in the parts.} and interdependence at a fundamental level.  We provide in reference \cite{QD} an illustration of this property: the Bell states are different bipartite states for which the reduced local states-also called reduced density matrices\footnote{It is commonly accepted that all the available information and knowledge that characterizes the state of a quantum system is contained in its density matrix. Quantum mechanics being a probabilistic theory, this knowledge is of statistical nature.}- are the same (as is shown in appendix of ref.\cite{QD}). In this approach, entanglement, non-locality and non-separability are manifestations of holism and Quantum Weirdness, to be opposed in our view to the classical, Cartesian, non-holistic approach in which the knowledge of the whole reduces to the knowledge of the parts and in which each system can be in principle be split into simple, separable, ultimate components. This raises the following question: is it legitimate to believe in the Cartesian paradigm (the description of the whole reduces to the description of its parts), when we know that the overwhelming majority of quantum systems are entangled?

In order to tackle similar questions, that are related to the so-called measurement problem \cite{measprob}, Zurek and coworkers developped in the 
framework of the decoherence approach \cite{decoherence} the idea that maybe,
 if we think classically, this is because during the evolution, our brain 
 selected in the external (supposedly quantum) world the islands of stability
  that correspond to the maximal quantum (Shannon-von Neumann) information \cite{reviewzurek}.
   These classical islands (environment induced or EIN\footnote{The EIN selection rule predicts that, once a system, its environment and their interaction are specified, the classical islands associated to the system are the states that diagonalize the reduced density matrix  of the system.  When the environment can be considered as an apparatus, the classical islands define the so-called pointer (or preferred) basis: {\it ``Pointer states can be defined as the ones which become
minimally entangled with the environment in the course of the evolution''} (quoted from Ref.\cite{abc}).} superselected \cite{EIN}) would correspond to the structures that our brain naturally 
   recognizes and identifies, and this would explain why the way we think is classical, for instance why we intuitively conceive the world as a collection of independent objects\footnote{At the level of Boolean logics this property corresponds to the principle of the excluded-third.}. Roughly speaking, the EIN selection principle expresses that, during the evolution, the classical islands that belong to the prefered basis or pointer basis are selected preferentially to any other basis. 
In the quantum Darwinist approach, the emergence of a classical world that obeys EINselection rules can be explained following two ways: 

A) these rules correspond to
maximal (Shannon-von Neumann) ''certainty'' or useful information (see footnote 17).

B) Zurek also invokes an argument of 
structural stability: superposition of states that would belong to such islands would 
be destroyed very quickly by the interaction with the environment which radiates irremediably the 
coherence into the environment \cite{decoherence}. This process is called the decoherence process and is very effective. 

This approach was baptised by Zurek under the name of Quantum Darwinism \cite{wikipedia}.

   In the reference \cite{QD}, we explained the meaning of quantum entanglement and described results that have relevance in the framework of the
    environment induced (EIN) superselection rules approach: 
    
    -(i) entanglement is the corollary of interaction (ref. \cite{QD} section (2.2), refs.\cite{zeit} and \cite{quote}); more precisely if an Hamiltonian of two systems $A$ and $B$ is such that all factorisable states remain factorisable (separable) during their evolution, then the interaction between them necessarily vanishes.
    
    -(ii)
     if we
    apply the criterion of maximal information to the simple situation during which
     two quantum particles interact through a position-dependent potential, in
      the non-relativistic regime then the classical islands are in one to one
       correspondence with the three classical paradigms elaborated by physicists before quantum mechanics existed; these are the droplet or diluted model, the test-particle and the material point approximations. In all these paradigms, the state of the particle ``factorizes'' from the environment, which is representative of the classical a priori conception of the universe.
       
       In the next section we shall study another physical system, that corroborates these results: two oscillators are coupled by an interaction that, essentially, consists of exchanging quanta (phonons are transfered between two oscillators $A$ and $B$). We apply the EIN superselection rule in this particular case and show that separable coherent states are EIN selected states, which means that they minimise the Shannon-von Neumann entropy of the reduced state.

      \section{Environment Induced Selection rule in the case of two coupled oscillators.}
Let us consider two coupled oscillators (or modes) that interact through the interaction Hamiltonian $H_{AB}=i\lambda \hbar (a^{\dagger}b-ab^{\dagger})$ (written in function of the creation and annihilation phonon operators of the oscillators $A$ and $B$).

The full Hamiltonian of the system constituted by $A$ and $B$ is then
\begin{equation}\label{ham}H=H_A+H_B+H_{AB}=\hbar\{\omega_A a^{\dagger}a+\omega_B b^{\dagger}b+i\lambda(a^{\dagger}b-ab^{\dagger})\},\end{equation} where $\omega_{A (B)}$ are the oscillator frequencies and $\lambda$ is the coupling constant between them.

 When the bipartite state $\Psi_{AB}$ is pure, the states that obey the EIN selection rule are such that the local or reduced Shannon-von Neumann entropy is minimal (equal to 0) which also means that the reduced states are pure. This occurs only when the full state $\Psi_{AB}$ is a product state, for the following reason. When $\Psi_{AB}$ is pure, it can be written (see section 5.5 of ref.\cite{QD}) in the Hilbert-Schmidt (biorthogonal) form\footnote{In the following expression, the sum can be considered as a sum-integral.}: $\ket{\Psi_{AB}}=\Sigma_{i}\alpha_i\ket{\psi_{A}}^i\ket{\tilde \psi_{B}}^i$ where $\ket{\psi_{A}}^i (\ket{\tilde \psi_{B}}^i)$ are orthonormal bases of the $A$ ($B$) Hilbert spaces. 
 
 The Shannon- von Neumann entropy is then equal to $-\Sigma_i| \alpha_i|^2log |\alpha_i|^2$, where the Schmidt weights $|\alpha_i|^2$ are positive real numbers comprised between 0 and 1 that obey the normalisation constraint $\Sigma_i| \alpha_i|^2=1.$ Now, the entropy is strictly positive unless all coefficients are either equal to 0 or equal to 1. By normalisation, only one Schmidt coefficient may be equal to 1 (say the $j$th one), all others being equal to 0. Then $\ket{\Psi_{AB}}=\ket{\psi_{A}}^j\ket{\tilde \psi_{B}}^j$ which is a product state.

Let us now assume that at time 0, a state is factorisable and let us impose that it remains so for all future times. According to the lemma proven in ref.\cite{zeit}, a pure product state remains a product state during the interaction if and only if,
during its evolution, the Hamiltonian never couples this product state to a product
state that is bi-orthogonal to it. 

The free parts of the Hamiltonian considered by us (eqn.\ref{ham}) do never couple a factorisable state $\ket{\Psi_{AB}(t)}$ to bi-orthogonal states, because they couple $\ket{\psi_{A}}^j\ket{\tilde \psi_{B}}^j$ to states of the form $(H_A\ket{\psi_{A}}^j)\ket{\tilde \psi_{B}}^j+\ket{\psi_{A}}^j(H_B\ket{\tilde \psi_{B}}^j)$ that are not bi-orthogonal to it. Now, the interaction potential couples the state $\ket{\Psi_{AB}(t)}$ to the space spanned by states of the form
\begin{eqnarray}H_{AB}\ket{\Psi_{AB}(t)}=i\hbar \lambda(a^{\dagger}b-ab^{\dagger})\ket{\psi_{A}(t)}^j\ket{\tilde \psi_{B}(t)}^j\nonumber\\=
i\hbar \lambda(a^{\dagger}\ket{\psi_{A}(t)}^j)(b\ket{\tilde \psi_{B}(t)}^j)-i\hbar \lambda(a\ket{\psi_{A}(t)}^j)(b^{\dagger}\ket{\tilde \psi_{B}(t)}^j).\end{eqnarray}

In the case that $\ket{\psi_{A}(t)}^j$ and $\ket{\tilde \psi_{B}(t)}^j$ are so-called coherent states which means that they are eigenstates of respectively $a$ and $b$, say for the (complex) eigenvalues $\mu_A(t)$ and $\mu_B(t)$ we get that

\begin{equation}H_{AB}\ket{\Psi_{AB}(t)}=
(a^{\dagger}\ket{\psi_{A}(t)}^j)(\mu_B(t)\ket{\tilde \psi_{B}(t)}^j)+(\mu_A(t)\ket{\psi_{A}(t)}^j)(b^{\dagger}\ket{\tilde \psi_{B}(t)}^j),\end{equation}

so that $H_{AB}\ket{\Psi_{AB}(t)}$ has no component in the space biorthogonal to $\ket{\Psi_{AB}(t)}$which is a sufficient condition, in virtue of the lemma, in order that the state $\Psi_{AB}(t)$ remains factorisable throughout the interaction between the two oscillators (modes). Now, let us check that the condition is valid at all times in the sense that if initially a state is a product of coherent states it remains a product of coherent states during its temporal evolution.

Let us denote $\ket{\mu_{A(B)}}$ the (coherent) eigenstate of the $a$ ($b$) operator for the eigenvalue $\mu_{A(B)}$. Let us try for $\ket{\Psi_{AB}(t)}$ the ansatz $\ket{\mu_A(t)} \ket{\mu_B(t)}$.

Its temporal evolution obeys Schr\"odinger's equation:
\begin{eqnarray}i\hbar{\partial \ket{\Psi_{AB}(t)}\over\partial t}=H\ket{\Psi_{AB}(t)}
=i\hbar({\partial \ket{\mu_{A}(t)}\over\partial t}\ket{\mu_{B}}+\ket{\mu_{A}(t)}{\partial \ket{\mu_{B}(t)}\over\partial t})\nonumber\\=\hbar\{((\omega_A \mu_{A}(t)+i\lambda\mu_{B}(t))a^{\dagger}\ket{\mu_{A}(t)} )\ket{\mu_{B}(t)}\nonumber\\+       
\ket{\mu_{A}(t)}(\omega_B\mu_{A}(t) -i\lambda \mu_{A}(t))(b^{\dagger}\ket{\mu_{B}(t)})\}.
\end{eqnarray}

It can be splitted into two local but not independent evolution equations:
\begin{equation}i\hbar{\partial \ket{\Psi_{A}(t)}\over\partial t}=H^A_{effective}\ket{\Psi_{A}(t)}\end{equation} and
\begin{equation}i\hbar{\partial \ket{\Psi_{B}(t)}\over\partial t}=H^B_{effective}\ket{\Psi_{B}(t)}\end{equation}
where we define the effective Hamiltonians as follows:

$H^A_{effective}=\hbar(\omega_A a^{\dagger}a+i\lambda \mu_{B}(t)a^{\dagger})$ and $H^B_{effective}=\hbar(\omega_B b^{\dagger}b-i\lambda\mu_{A}(t)b^{\dagger})$.

Now, Glauber has shown in 1966 \cite{glau66} that for a large class of ''local''\footnote{We mean here single-mode.} Hamiltonians  coherent states remain coherent states throughout evolution (a generalization of an observation originally \cite{schrodcoherent} made by Schr\"odinger\footnote{This property, that coherent states remain coherent was a criterion of classicality according to Glauber. Besides, it can also be shown that the coherent oscillator states minimize Heisenberg uncertainties. Essentially this is so because they can be obtained by letting act displacement operators onto the vacuum state, the ground state of an harmonic potential, which minimizes position-momentum complementarity and because displacement operators preserve position and momentum uncertainties. Glauber's result corroborates Schr\"odinger observation according to which a translated ground state will coherently oscillate around the origin, as would do its classical counterpart...}). 

This class consists of ''local'' Hamiltonians of which the commutator with the ''local'' annihilation operator (say $a$) is still a function of $a$ and time alone (in other words this commutator does not contain the creation operator $a^{\dagger}$ or any of its powers). It is straightforward to check that the effective Hamiltonians $H^A_{effective}$ and $H^B_{effective}$ belong to this class of coherence-preserving Hamiltonians. Actually the highest power to which creation operators $a^{\dagger}$ and/or $b^{\dagger}$appear in the expression of those effective Hamiltonians is one, which is sufficient for guaranteeing coherence-preservation. Now, these effective Hamiltonians are not strictly speaking local Hamiltonians but the ''non-local'' coupling (we mean here the interaction between the two oscillators) is expressed in last resort through the explicit time-dependence of the effective Hamiltonians and does not lead, in the example studied here, to the appearance of quadratic and higher powers of the creation operator(s). 

This explains why products of coherent states remain products of coherent states despite of the coupling.
\section{About a Quantum Signature in Biological Activity.}
\subsection{Recent Observations gathered about Light Harvesting Bacteria.}
As we discussed before, it could be relevant to try to find the trace of a quantum signature at the level of primitive living organisms. Now, recent experiments have shown that it is very likely that certain bacteria exhibit some form of quantum coherence. These bacteria are called Light Harvesting Bacteria. They live at the bottom of deep lakes or of the sea, in an energetically poor environment and they literally ''eat photons''. They attracted the attention of biophysicists during several decades among others because they seemingly manage to transform photonic energy into chemical energy (ATP) with an extremely high efficiency (97 \%). Although this process is well understood from a biochemical point of view, the deep reason of this very high efficiency is not yet fully elucidated (to compare, manufactured solar cells exhibit an efficiency at most of the order of 20 \%).

  During the last decade, several clues suggested that the light absorption process that takes place at the level of the receptors of these bacteria could 
involve coherent collective excitations (excitons-polarons) and that quantum mechanics could play a fundamental role in the process. In a 2007 Nature paper \cite{fleming}, entitled \textit{``Evidence for wavelike energy 
transfer through quantum coherence in photosynthetic system''}, Fleming and his collaborators report the detection of \textit{''quantum beating''} signals, coherent electronic oscillations in both donor and acceptor molecules, generated by light-induced energy excitations. In their paper one could read the following sentences:
 
\begin{quote}
 ``We have obtained the first evidence that remarkably long-lived wavelike electronic coherence plays an important part in energy transfer processes during photosynthesis''

  ''The classical hopping description of the energy transfer process is both inadequate and inaccurate, it gives the wrong picture of how the process actually works, and misses a crucial aspect of the reason for the wonderful efficiency'' 
  
  ``The remarkably long-lived wavelike electronic coherence can 
 explain the extreme efficiency of the energy transfer because it enables the system to 
 simultaneously sample all the potential energy pathways and choose the most efficient ones'' 
 \end{quote}
Although in our eyes, the observations are not conclusive (for instance it is not clear whether the observations reveal quantum coherence or merely classical coherence), they are stimulating and serve our purposes. Interestingly, our very simple model of two coupled quantum oscillators perfectly fits the observations of Fleming \textit{et al.}. It is also a valid model in the present case because at the rotating wave approximation \cite{milburn}, the coupling between a particular electro-magnetic mode and an oscillating dipole is precisely described by a Hamiltonian that obeys equation \ref{ham} \footnote{It is worth noting that energy conservation imposes that both modes oscillate at equal frequencies. This corresponds to a situation where the oscillating dipole and the electro-magnetic mode are resonant.}.

Now, a coherent beating is altogether the manifestation of classical coherence, or quantum quasi-classical coherence as shows the discussion of the previous section so that a long coherence time cannot be considered as a conclusive proof that quantum coherence is present.

Obviously, in order to clearly distinguish between classical and quantum coherence, the most convincing experimental argument would consist of showing explicitly that entanglement is present during the interaction between a living organism and electro-magnetic radiation\footnote{During the revision process of this paper, we discovered another proposal aiming at revealing the presence of entanglement during biological processes, in this case, during the absorption of a photon by light-harvesting bacteria \cite{fleming2}.}. This is because correlations exhibited by entangled systems make it possible to violate Bell inequalities \cite{34}, which is very difficult and sometimes impossible to realize with classical correlations, in agreement with the aforementioned quotation of Schr\"odinger according to who entanglement is the characteristic trait of quantum mechanics,
``the one that enforces its entire departure from classical
lines of thought'' \cite{erwin}.

 In the next section we propose an experimental scheme that in principle could allow us to reveal, if it exists, the presence of entanglement during a process called ultraweak long range delayed biophotonic emission. The typical intensities of this ultraweak biophotonic emission are so weak that it constitutes a good candidate for trying to find a quantum signature. This is because the signal is detected photon by photon, which corresponds intuitively to the realm of Quantum Optics and Quantum Information theories.    
 
 It is then relevant to investigate whether the techniques tools and concepts developed in the framework of those disciplines could help us to learn something new about biophotonic emission\footnote{This is relevant but maybe not so reasonable a priori, because it is usually considered that at the scale that characterizes living organisms, even the smallest ones like individual cells, decoherence will be very efficient. Of course, it is possible to realize mesoscopic physical systems that exhibit a high level of coherence like SQUIDS \cite{squids} but these systems cannot be realized in the conditions (temperature, thermodynamical phase and so on) that characterize living systems. Now, it is suspected that in order to explain the extreme efficiency of light harvesting bacteria, a collective (quantum) behavior \cite{collective} ought to be present (excitons). Considered so, it is relevant at least to try to reveal the existence of quantum coherence and entanglement at the level of biological processes. Interferometric-Schr\"odinger cats-like tests have revealed that quantum coherence is still there at the level of macro-molecules such as porphyrin, a component of hemoglobyne \cite{zeilinger}. It is an open question however to know whether living organisms would survive such tests \cite{cirac}.}. 
\subsection{Quest of a Quantum Signature in Biophotonic Emission.}
\subsubsection{Ultraweak long range delayed biophotonic emission.} There exists an impressive literature  \cite{biowiki,gurwitsch,strehler,popp81} about so-called delayed bio-luminescence
 (or biophotons). This consists of a long-term afterglow that appears when 
 living cells, after having been exposed to an intense flash of light, reemit a part of this energy after a 
  non-negligible time-delay of the order of one hour. This process occurs in a large
   range of wavelenghts (200 to 800 nm) and with intensities ranging 
  from a few up to some hundred thousand photons by centimeter squared-second; it is twenty
   orders of magnitude smaller than the common (short-term) fluorescence
   or phosphorescence and 10 orders of magnitude larger than thermal radiation at ambient temperature.  This phenomenon is seemingly \cite{[PY02],[PCHYY02]} characterized by a high degree of coherence\footnote{In ref.\cite{[PY02]} a model is also developed with the aim of associating coherent states to biophotons. It is based on an explicit solution derived by Mehta and Sudarshan \cite{sudar} for the (local) time evolution of a coherent state of which the Hamiltonian belongs to the class of coherence-preserving Hamiltonians considered by Glauber in ref.\cite{glau66}. As the authors have shown, this approach leads to the prediction of a hyperbolic-like relaxation function that fits well the observed data.}; for instance the typical decay curves of the biophotonic intensities exhibit oscillating in time modulations \cite{[PY02]}, somewhat reminiscent of Rabi oscillations (a typical signature of quantum coherence, that was for instance observed by high energy and particle physicists in the study of so-called kaonic oscillation and regeneration  \cite{kaon}).
   
   Now, classical damped oscillators and coupled chemical reactions also exhibit such oscillations \cite{brussellator,kaonreal,hamamatsu}, and it is uneasy to decide whether their origin should be, in last resort, attributed to quantum coherence, classical coherence or merely to some synchronized biochemical activity characterized by antagonist coupled chemical reactions.

   In the reference  \cite{[PY02]}, the authors emit the hypothesis that the biophotonic signal exhibits quantum coherence in the sense that the biophotons would be in a quantum coherent state. Their hypothesis was motivated by the fact that the statistics of the detection clicks accumulated in a given temporal window are Poisson-distributed. 
   
   According to us this hypothesis has not yet been convincingly validated on experimental grounds because it is not easy to differentiate a quantum coherent signal from other stochastic signals simply by considering the statistical data obtained from a single photo-detector.  
   
   We agree that in the case of a coherent source, it is well so that the temporal distribution of time delays is exponential while the corresponding populations (accumulated over a fixed temporal window) are Poisson-distributed. 
   
   Now, the problem is that plenty of other stochastic processes exist in Nature that do not exhibit any kind of coherence but lead to the same distributions. The distribution of decay-times of a sample of radio-active U-238 isotopes exhibits for instance exactly the same properties. It is therefore not easy to establish definitively quantum coherence by such methods and we consider that extra-tests are necessary in order to confirm the quantum nature of biophotons and/or the fact that they would be produced in quantum coherent states.

\subsubsection{Two-detector statistics: bunching and anti-bunching.} A more sophisticated way to study the properties exhibited by biophotons is to consider the second order statistics that we obtain when we measure correlations exhibited by two detectors \cite{hbt}.  Roughly speaking, when the second detector is more (less) likely to click directly after a click in the first one, the distribution is said to be bunched (anti-bunched). When the probability to click per unit of time is independent on time which constitutes an intermediate situation between {\it bunching} and {\it anti-bunching}, the statistics is Poissonian.
   
    In the reference \cite{[PCHYY02]}, the authors emit the hypothesis according to which the biophotons are squeezed in the sense that their coincidence 
   statistics exhibits experimental departures from the Poisson statistics (bunching or antibunching-which is a signature of non-classical light \cite{sudarshan}). 
   
   As is well-known, bunching (which reveals the tendency of photons to arrive together in both detectors) is a property that is also exhibited by classical incoherent light (for instance thermal light), so it is not a convincing ``quantum signature''. 
   
   Anti-bunching on the contrary has been observed so far only with individual bosonic sources (for instance single photon sources like trapped atoms) and reveals that the source acts as a whole \cite{antibunching}  so that the presence of anti-bunching in biophotonic emission would be very intriguing and challenging by itself \cite{sudarshan}. Now, it is notoriously known that biological systems are charcterized by a high versatility and one could wonder whether the observation of anti-bunching reported in the reference \cite{[PCHYY02]} could not be accidental. In order to firmly establish statistically the presence of anti-bunching, it is certainly worth repeating this kind of experiments. The main problem is that, as a consequence of the law of large numbers, the quantity of data that is required in order to establish anti-bunching in a statistically significant manner is large \cite{baudon} although the biophotonic emission on which the data is based is ultraweak, so that the collection time becomes huge which menaces the reproducibility of the experiment. 
   
   From our point of view, in order to track the entanglement signature in the biophotonic signal, the most promising strategy is to realize a more refined experimental study of the two-detector statistics of biophotonic emission as we shall discuss in the next section.
   
   \subsubsection{Quantum tomography of biophotonic emission: Measuring the degree of entanglement exhibited by bio-photons.} The goal of quantum tomography is to characterize the density matrix of an a priori unknown signal  \cite{[BAMOMBRH],[TD06a],[DLLK07]}. In principle it makes it possible to perform a precise and complete experimental study of the properties of biophotons (correlations, entanglement and
    so on). This in turn could allow us to test whether the biophotonic correlations allow us to violate well-chosen Bell's inequalities, which establishes a clear border line between classical and quantum systems, and is at the same time a measure of the degree of entanglement and a clear signature of the non-classicality of a signal. This constitutes our experimental proposal: to realize full quantum tomography of the biophotonic light states in order to investigate their non-classical features. 
Once tomography is performed, we are able to estimate the coefficients of the corresponding density matrix  \cite{[BAMOMBRH],[TD06a],[DLLK07]}.
Then, it is possible in principle to estimate the degrees of correlation, entanglement and separability of the photonic signal (for instance thanks to entanglement witnesses \cite{witness}).

Entanglement is a versatile property that can characterize all types of quantum degrees of freedom (discrete, continuous), from a same system (for instance entanglement between internal degrees of freedom (like spin or electronic energy level) and external degrees of freedom of a same atom (like position or momentum)), or from different systems\footnote{Actually, there exists no universal mathematical measure of entanglement about which everybody agrees, excepted in the most simple cases (for instance for bipartite systems in a pure state, or for arbitrary two-qubit states \cite{witness}). As a consequence, the study of entanglement became a discipline of mathematical physics in itself, and it is clearly out of the scope of our paper to review this huge subject.
  }.

Therefore, we are free to attempt to reveal the presence of entanglement in the biophotonic emission in various degrees of freedom such as for instance frequency, or polarisation, or momentum  and so on. 

Moreover, there exist several ways to choose the {\it correlators}, depending on (i) our experimental setting, (ii) the quantum state of the system under study and (iii) which type of Bell inequalities we desire to violate, so that the number of possible experimental strategies is virtually unlimited.

 It is instructive nevertheless to focus on the most simple experimental proposal in which we aim at revealing entanglement between polarisations of biophotons collected along two different directions, which will also allow us to compare our experimental proposal with the famous Orsay experiment\cite{orsay} that is recognised to be the first convincing experimental evidence of the violation of local realism (by pairs of photons entangled in polarisation).
\subsubsection{Quantum tomography of biophotonic polarisations: Orsay experiment with biophotons.}
In order to estimate the full polarisation state of a pair of photons, there exist two main approaches that are outlined in the section 3 of ref. \cite{[TD06a]}. In the first, most conventional, approach, the polarisation of each photon is measured locally in 3 mutually unbiased bases, and by estimating the 36 joint-probabilities that are collected so, one is able to determine all the coefficients of the 4 times 4 density matrix that characterizes the full polarisation state of the pair. In the second one, a Symmetric Informationally Complete POVM is realized separately on each photon, and once again full tomography can be obtained after having estimated the (16) joint-probabilities assigned to the local detectors (4 at each side) implied in the POVM tomographic procedure.

Once the two qubit density matrix of the polarisation state of the pair is known, it is sufficient to compute the concurrence (which is a well-defined measure of entanglement in the case of arbitrary two-qubit systems \cite{concurrence1}) in order to estimate its degree of entanglement.

It is instructive to compare this experiment with the experiment realised by Aspect and coworkers in Orsay in 1980. In that experiment pairs of photon were emitted during a cascade process in such a way that they were entangled in polarisation (the pairs were actually prepared in a Bell-singlet state that is maximally entangled). Then the correlations between various bases of polarisation were measured, among others along the bases that maximised the violation of Bell-like inequalities (two at each side were sufficient for doing so). It was not necessary during that experiment to perform tomography of the two-photon state because this state was tailored on purpose (exploiting spectroscopic properties of the source and resorting to general conservation principles, of angular momentum in that case, that guaranteed the presence of entanglement).

The Orsay experiment \cite{orsay} flirted with the limit of acceptable signal-noise ratio, due to the fact that, in order to avoid the locality loophole, the distance between source and detectors was taken as long as possible.

In our case we ignore whether the biophotons are entangled or not, because we are not able to trigger their emission and their quantum state will remain unknown to us before we measure it experimentally. Nevertheless, due to the fact that long delayed biofluorescence is a very weak intensity process, photons are detected one by one, and, as in Orsay experiment one can impose coincidence windows in order to be able to discriminate the rates of simultaneous pair emissions and of fortitious joint-detections.

Conceptually, our proposal is very similar to the original Orsay experiment: a source emits photons that are filtered in polarisation and detected in two spatially separated photo-detectors. The correlations exhibited by the detectors could possibly allow us to reveal the presence of entanglement and to violate well-chosen Bell inequalities.

\section{Discussion and conclusion.} 

The picture of the world that modern physics offers lies far away from the idyllic image of a garden of Eden. According to our present picture of the world at macro-scales, the Universe results from a cosmic ''explosion'', our planet is lost in a huge emptiness, around a star that will certainly collapse one day and it could be that physical constants themselves are not constant throughout time. Even subatomic particles are not stable: they transform into each other, their behavior is intrinsically unpredictible and because of quantum entanglement it is most often a non-sense to consider them as isolated objects. A machinery that would possess such properties would constitute a real nightmare for most engineers and technicians.

Besides, scientific disciplines such as chemistry, biology, medicine, psychology, are heavily compartimented and nearly don't talk to each other. Also here the vision of a garden of Eden is seriously compromised: we learn from biology that monkeys are our cousins, and from psychology that our behaviour itself is most often out of control: we are quite closer to Mr Hyde than to Dr Jekill, obviously.

In view of all this it is thus legitimate to doubt about the solidity of our representation of the world and it is interesting to consider carefully the hypothesis according to which our world-view is itself a dynamical process, submitted for instance to mechanisms of natural selection. This hypothesis was at the core of the present paper\footnote{One could wonder whether this paper deserves to be published in a special issue of Studia Logica devoted to Contributions of Logic to the Foundations of Physics or rather in a ``regular'' physical journal. In our view, Logic and Mathematics are themselves empirical sciences to some extent, which is consistent with the thesis developed here according to which our worldview has been modeled by the environment during a long evolution process. Considered so, any experimental quest of non-classicality can be considered as a contribution to Logic that is likely to interest the audience of Studia that consists mainly of logicians and philosophers. We did our best to avoid to be too technical and esoteric in our presentation. } and of a previous one \cite{QD} where we investigated the idea that what we call classical reality could result from a long selection process.

The criterion according to which certain perception channels and/or representations of the world would be preferentially selected is that they maximize the amount of (useful) {\bf information}\footnote{There exist several notions of information and Shannon's type of quantitative measures are only one way to characterize it. It is important to
note that this concept, although fundamental, is difficult to use, and not always
defined in the same manner, as illustrated by the comment and response that can be found in ref.\cite{EDU},  pp154 and 157. We also explain in our response (p157, same reference) what we mean by {\it useful} information.} that they allow us to extract from the external world. In last resort it is this criterion that would explain the emergence of ``classicality''.

In the present paper we did our best to go beyond sterile academic discussions and to propose simple experiments aimed at investigating whether the presence of quantum coherence and quantum weirdness could be revealed at the level of basic biological activities. We emitted the hypothesis that,  in the ultraweak intensity regime, biophotonic emission could possibly exhibit non-classical features such as entanglement and proposed to test this possibility by performing quantum tomography of biophotonic states of light.

The ball is now in the camp of experimentalists. It is our hope that, as soon as possible, Nature will have the last word in this discussion...

\subsection*{Acknowledgements.} 
The author acknowledges support from the ICT Impulse Program of the Brussels Capital Region (Project Cryptasc) and the European Project Photonics4Life, and also from
  the IUAP program of the Belgian Science Policy Office (grant IAP P6-10 {\it photonics@be}) and the Solvay Institutes for
 Physics and Chemistry.

\end{document}